\documentclass[epj]{svjour}
\usepackage{graphics}
\usepackage{graphicx}
\usepackage{amsmath}
\usepackage{braket}
\usepackage{subfig}
\begin{document}
\title{Quantum Dynamics of a Two-level System Near a Gold Nanoparticle}
\author{Somayeh M.A. Mirzaee\inst{*},and Chitra Rangan
}                     
%
%
\institute{Department of Physics, University of Windsor, Windsor ON, Canada \\ \inst{*} Present affiliation: Department of Physics and Astronomy, Queen’s University, Kingston ON, Canada}
\date{Received: date / Revised version: date}
%
\abstract{
The ability to prepare desired states without allowing time for the atomic system to spontaneously decay to the ground state is limited by the fact that Hamiltonian controls cannot affect the purity of the state in question. In this paper, we discuss how a continuous off-resonant electromagnetic wave can be used to prepare a desired final quantum state. With the aid of plasmonic surface-enhancement, by placing a two-level atomic system near a gold nanoparticle, preparation is accomplished in a reduced period of time and with lower electric field intensity while the system purity is improved when compared to an isolated system. 
\PACS{
      {PACS-key}{discribing text of that key}   \and
      {PACS-key}{discribing text of that key}
     } 
} 
\maketitle
\section{Introduction}
\label{intro}
The issue of quantum computation due to its ability to perform certain types of calculations much faster than classical computers has attracted much attention recently. The basic concepts of quantum computation are quantum operations (gates) on quantum bits (qubits) and registers (arrays of qubits). A qubit can be a two-level system which can be prepared in arbitrary superposition of its two eigenstates, usually denoted as $\ket{0}$ and $\ket{1}$ . In that sense quantum state engineering is needed to control preparation and manipulation of these quantum states. 
In quantum information systems, decoherence, due to the interaction of the quantum state with the environment, represents a fundamental issue that must always be addressed as it results in a leakage of system information to the environment \cite{1a}. Therefore it is desirable to have quantum states that can be prepared and purified in a short period of time to avoid this loss of information. 
A great deal of recent research has been devoted to proposing solutions, such as entanglement purification\cite{2a}, that allow for the preservation of this purity of states \cite{3a,4a,5a,6a,7a}. These methods include quantum error-correcting codes\cite{8a,9a}, decoherence-free subspaces\cite{10a,11a,12a}, strategies based on feedback or stochastic control \cite{13a,14a,15a,16a} as well as dynamical decoupling \cite{17a,18a,19a}. These methods all share a common strategy: they fight against de-coherence from the environment by decoupling the quantum system from the environment. We show here that dissipation, if engineered, can have exactly the opposite effect: it can be a well-developed resource for universal quantum computation. To overcome noise, purification techniques, which generate qubits with higher purities from qubits with lower purities, have been proposed\cite{20a}.
As the purification dynamics of quantum systems are quantfied by the average purification rate and the mean time needed to reach the required level of purity\cite{21a}, designing systems that optimize these quantities are highly desirable. 
This problem of finding effective purification schemes for quantum states is difficult due to fact that the purity of a quantum system cannot be changed by Hamiltonian controls alone\cite{22a} and require some other effects in order to modify the purity of the states in question. One such effect that can be used to modify the purity of states is through the inclusion of spontaneous decay. However, using spontaneous decay to purify quantum states requires weak driving fields and long time scales in order to reach a desired state of purity. 
In this work, we show that the presence of a gold nanoparticle (GNP) near individual qubits can be used to enhance the purification rates of these qubits when using an incident electric field as in laser cooling. This enhancement in the purification rate arises due to the modification of the total decay rate of a two-level quantum system coupled to a nonlocal plasmonic particle \cite{23aa}, or to a localized surface plasmon mode\cite{23a,24a} of a GNP plus local electric field enhancement. Moreover, we will prove that the proximate GNP can use to preserve the purity of the quantum information (QI) carriers in presence of decoherence processes, which can be used in transmission process.

\section{System Model}
A schematic illustration of our model is given in figure 1. We model the qubit as a molecular two-level system with states $\ket{g}$ and $\ket{e}$. This qubit is placed a specific distance $’d’$ from the surface of a GNP. This quantum system is driven by an electromagnetic wave that is incident on the entire system. 
The Hamiltonian of the two-level quantum system that is driven by an electromagnetic wave, with the field-matter interaction of the system treated in the electric-dipole approximation described by $\hat{H} = \hat{H}_a + \vec{\mu} \cdot \vec{E}(t)$ , where $H_a$ is the Hamiltonian of a two-level system, $\vec{\mu}$ is the transition dipole moment of the system and E(t) is electric field of the wave interacting with the system. This electric field magnitude varies in time as $E(t) = E_L cos(\omega_L t)$ where $\omega_L $ is the frequency of the incident wave.  
Making the rotating wave approximation (RWA), the matrix form of the Hamiltonian can be written as: 
\begin{equation}		
	   H_{RWA} =  \left ( \begin{matrix} 0 & \frac{\hbar\Omega_{ge}}{2} \\ \frac{\hbar\Omega_{ge}^*}{2} & -\hbar \bigtriangleup \end{matrix} \right ) ;
		\end{equation}

\label{sec:1}

\begin{figure}
\resizebox{0.5\textwidth}{!}{%
  \includegraphics{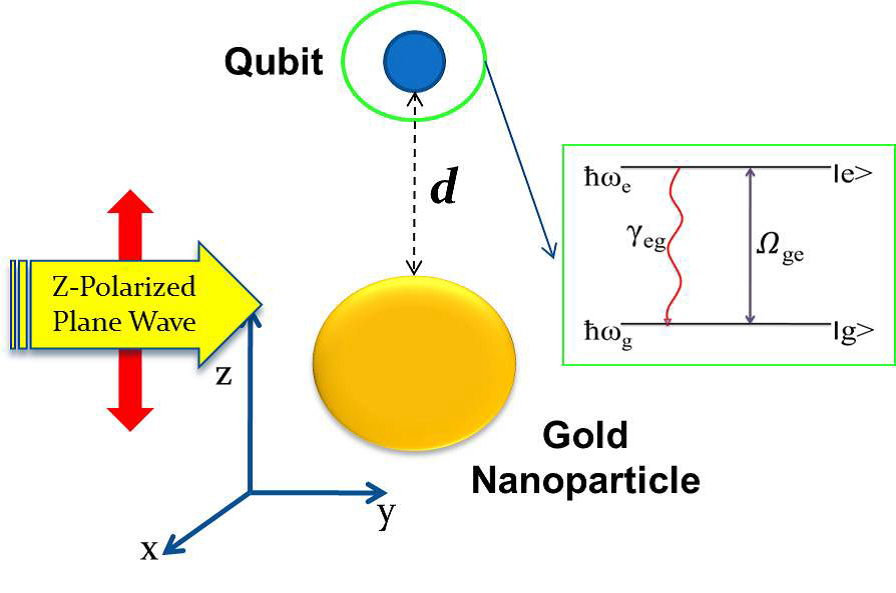}
}
\caption{Schematic illustration of the model system. This system consists of a qubit placed near a GNP and the entire system is under the influence of a classical electric field.}
\label{fig:1}       
\end{figure}

where  $\bigtriangleup$ represents the detuning between the frequency of the incident field and frequency of the state transition. The Rabi frequency, $\Omega_{ge} = \frac{\mu_{ge} M_L E_L}{\hbar}$, depends on the amplitude of the local electric field ($M_LE_L$), where the incident electric field is modiﬁed by the interaction of the electromagnetic wave with the GNP. This interaction is reflected in the inclusion of the frequency-dependent field enhancement term, $M_L$, which affectively increases or decreases the electric field intensity experienced by the two-level system. 
In order to study the time-dependent response of the qubit to both the environment and the incident electromagnetic wave, we use a density matrix representation of the qubit’s state. For our two-level system this density matrix is of the form: 

\begin{equation}
\rho = \left ( \begin{matrix} \rho_{gg} & \rho_{ge} \\ \rho_{eg} & \rho_{ee} \end{matrix} \right ) ;
\end{equation}
	This density matrix evolves in time under the LVN equation which takes the form:
\begin{equation}		
\dot{\rho} = - \frac{i}{\hbar} [H_{RWA},\rho] - L ;
\end{equation}		
	In this evolution equation, the Lindblad term, L, models the decoherence in the system. This term is linear in the state density operator and is of the form:
\begin{equation}		
	\hat{L} =\sum_d{\frac{ M_d\gamma_{eg}}{2} (\sigma_d^\dagger \sigma_d \rho + \rho\sigma_d^\dagger \sigma_d - 2 \sigma_d \rho \sigma_d^\dagger) } ;
\end{equation}
In this equation, $\sigma_d$ are the Lindblad operators and we assume that the only decoherence mechanism present is spontaneous emission.  Therefore $\gamma_{eg}$ represents the decay rate from the excited states to the ground state and we take $\sigma_d^\dagger = \ket{g}\bra{e}$.  The factor $M_d$ represents a modiﬁcation of the decay rates that arises from an interaction between qubit and the environment \cite{23a}.
 In this work we quantify the purity of a state as $Purity = Tr (\rho^2)$ \cite{25a}.  Under this description, states with a higher purity are closer to being pure states. It is important to note that if only Hamiltonian controls are applied to the system this state purity cannot change and therefore decay must be included to change the purity of measured states \cite{22a,23a,24a,25a}. Due to the increased decay rate, brought on by proximity to the GNP, this change in purity occurs on a shorter timescale. The enhancement quantities $M_L$ and $M_d$ are reflective of the effect of the GNP and next we show how these quantities are computed.

\section{Effect of a GNP on a Proximate Two-level System}
\label{sec:2}

Light incident onto a GNP can excite localized surface plasmon resonances that enhance the electric field around the nanoparticle \cite{23a}. In addition to these resonances, metal particles amplify the decay rates of excited states in nearby molecules by offering non-radiative decay channels and by coupling radiative emissions to a localized surface plasmon mode \cite{23a,26aa}. These interactions can greatly affect the response of a qubit system to an incident electric field. 
In order to calculate the field enhancement and polarization around a GNP, we solve Maxwell’s equations for the interaction of the incident electromagnetic field with the GNP. For a point GNP, this calculation can be done analytically \cite{27a}. In order to extend our formalism to GNPs of irregular size and configuration we solve Maxwell’s equations numerically \cite{28a}. We do this by performing a three-dimensional finite-Difference Time Domain (FDTD) calculation using the FDTD Solutions software package from Lumerical Solutions Inc. This also allows us to determine the vector components of the electromagnetic field at different locations as opposed to an average field enhancement. The dielectric function of gold was chosen to be “Au (gold) - Johnson and Christy” from the material database.  
Figure 2 presents the calculated electric field components relative the incident electric field amplitude for the case of a GNP of radius 20 nm that is illuminated by a z-polarized plane wave of $\lambda = 542 nm$ propagating along the z direction; this wavelength corresponds to the maximum electromagnetic enhancement around the GNP. We have seen that, as the particle diameter is no longer small compared to excitation wavelength, plasmonic modes of higher order than the dipolar mode becomes important and contribute to the electric field distribution. We assumed that the parallel orientation of dipole moment relative to the nanoparticle surface is excited only by the tangential component of electric field, while only the radial component contributes to the dipole with the perpendicular orientation. Therefore, unlike the literatures that focus on total electric field enhancement, we have calculated the electric field components around the GNP. These electric field components can be used to determine the electric field enhancement $M_L$ at any given point relative to the GNP. We will place the qubit away from the GNP along its z axis as this area represents the strongest area of electric field enhancement. We will also assume that the qubit polarization is along the z direction as the field in the y direction in ten times weaker than the field in the z direction. 

\begin{figure*}
 \includegraphics{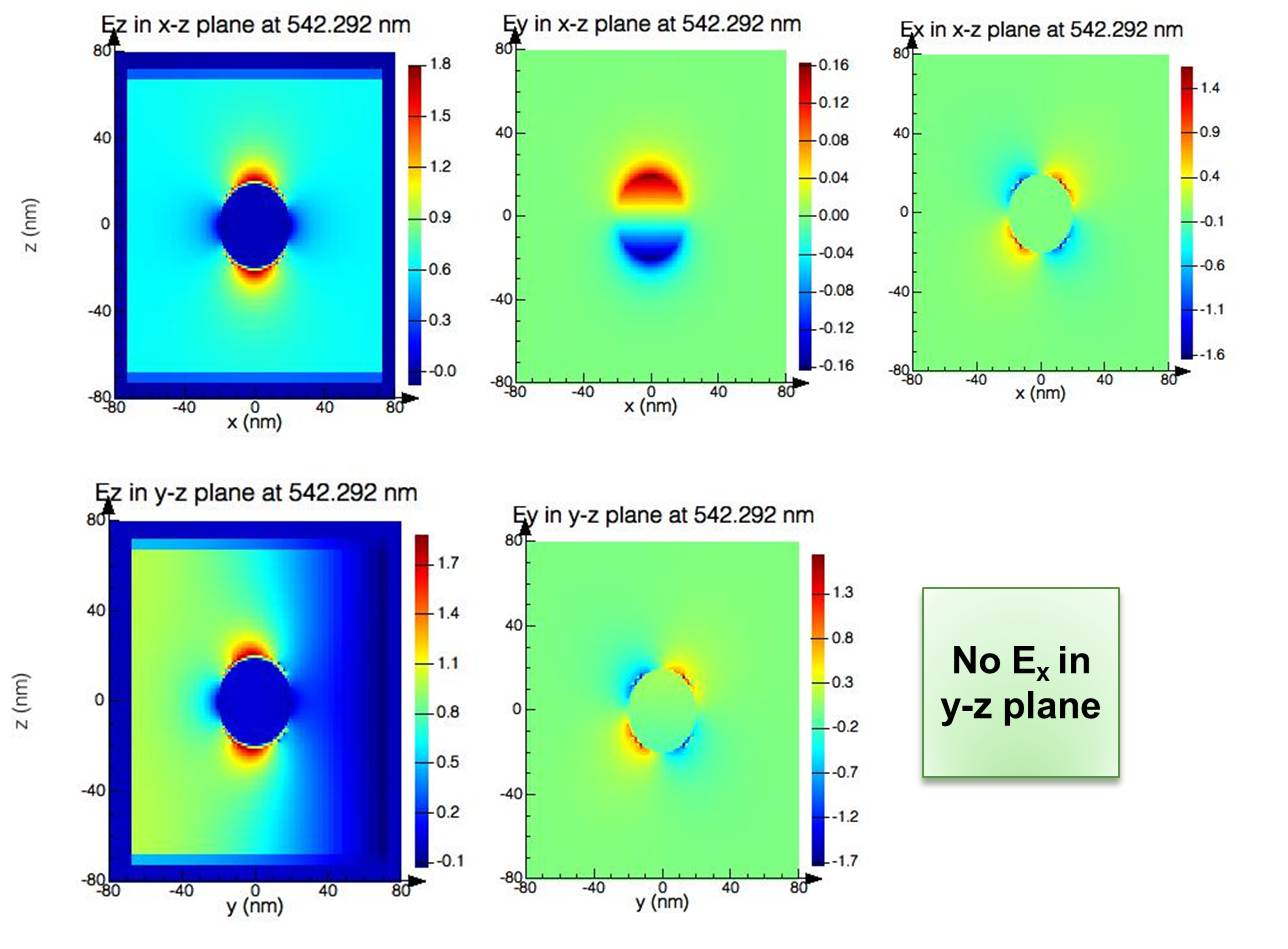}
\vspace*{1cm}       
\caption{Calculated electric field components, relative to the incident electricfield, for the case of a GNP of radius 20 nm that is illuminated by a z-polarized plane wave of λ = 542 nm propagating along the y direction }
\label{fig:2}       
\end{figure*}

This numerical simulation of Maxwell’s equations can also be used to calculate the enhancement of the decay of the excited state ($M_d$). To calculate the decay rate enhancement factor in the presence of a gold nanoparticle, the same model as Gresten-Nitzan model\cite{29a} was employed. The system of radiating dipole interacting with an isolated GNP is solved using the finite-difference time-domain (FDTD) method. The quantum system is simulated be a point radiating dipole source in the near fiield of the nanoparticle. The parallel and perpendicular orientations of the radiating dipole moment relative to the nanoparticle surface are examined and by placing these model dipoles at varying distances from the GNP and in various orientations with respect to the GNP surface, we can examine the effect of GNP-qubit separation on the decay rates of the qubit system. The analysis of the FDTD results relies on the fact that, for an atomic dipole transition, the normalized quantum mechanical decay rate can be related to the normalized classical power radiated by the dipole. In the presence of absorption, this classical approach allows one to calculate separately the radiative decay rate (proportional to the far field radiated power) and the nonradiative decay rate (proportional to the power absorbed by the environment) \cite{30a}.
In the literature of surface-enhanced fluorescence, it is normally assumed that the dipole placed a distance away from a GNP experiences enhanced spontaneous emission \cite{23a}. Figure 3 shows the radiative and nonradiative decay rate modiﬁcation experienced by the qubit system at various distances from the 20 nm radius GNP for the qubit orientations both parallel and perpendicular to the GNP surface. From figures 2 and 3 we can see that the interaction between a qubit and the GNP is not only distance dependent; it is effected by both the orientation of the qubit as well as the location of the qubit around the GNP. By placing our qubit system at different locations, we are given a degree of control over the decay rates and local fields experienced by the GNP.

\section{Atomic System Preparation in Desired State}
\label{sec:3}

In order to prepare a qubit in a desired state, we investigate the dynamics of a two-level system when it is excited by a classical electric field in a dissipative environment. Our idea is to engineer those couplings, so that the environments drive the system to a desired final state. Equation 3 describes the evolution of the state densities over time. By rearranging Equation 3 into differential equations for both the ground state populations and coherences, we find that at steady state ($\dot{\rho}=0$ ) conditions, the final state populations of the system are :
\begin{equation}
	\rho_{gg} =1 - \frac{\frac{\Omega_{ge}^2}{4}}{\bigtriangleup^2+\frac{\gamma^2}{4}+\frac{\Omega_{ge}^2}{2}} ;
	\end{equation} 
	\begin{equation}
	\rho_{ge} = \frac{\frac{1}{2}\bigtriangleup \Omega_{ge}+i\frac{1}{4}\Omega_{ge}\gamma}{\bigtriangleup^2+\frac{\gamma^2}{4}+\frac{\Omega_{ge}^2}{2}} ;
	\end{equation} 
	
	Where $\gamma$ represents the total decay rate. These equations indicate which final state the system will end up in, \emph{regardless of initial conditions}, as long as the system evolves to reach a steady state. In order to illustrate this, we have evolved a qubit at a distance of 5 nm from a GNP for a variety of initial conditions. For this system, the energy separation is 2.38eV and the dipole moment is chosen to be $\mu_{ge}=1e\AA $, where $e$ is the charge of an electron. We assume that the frequency of our incident electric field is chosen to yield a photon energy of 2.29eV which leaves us with a detuning  $\bigtriangleup= 0.09eV$. Figure 4 shows this system evolving, under a chosen laser intensity, to a desired final ground state population $\rho_{gg} = 0.75$, for a variety of initial mixed states. Each state is initialized with no coherence, $\rho_{ge} = 0$ and evolves to the same final state in the same amount of time without any active control. This independence of the final state on initial conditions indicates that a continuous electromagnetic wave can be used to purify a mixed quantum state if the purity of the initial state is lower than the final state that will be reached. It also places a fundamental limit on the amount of purification that can be done on the system. 
This differs from pulsed excitation schemes in that the purity can be changed and the initial conditions do not need to be known; avoiding the need to allow the system to decay completely to the ground state. 
In addition the GNP also helps to greatly reduce the required electromagnetic intensity required to reach the desired final state. As both field enhancement and decay enhancement are dependent on the distance, we can examine this effect by placing the qubit closer to the GNP. Figure 5 shows the state dynamics and purity by placing the qubit at a distance $2\hat{z}  nm$ from the GNP. Figure 6 shows, this placement reduces the required electromagnetic field due to the electromagnetic field enhancement effect versus the required electric field for the bare qubit, which is around $5\times10^9 V/m$.

\begin{figure*}
 \includegraphics{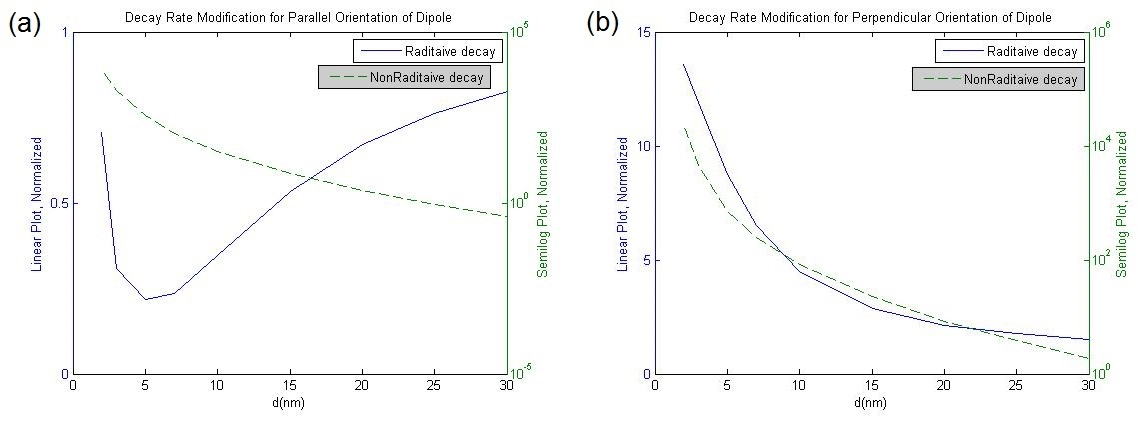}
\vspace*{1cm}       
\caption{Radiative and nonradiative decay rate modiﬁcation of a radiating point-dipole around a GNP with r = 20 nm at various distances from the GNP surface. (a) represents when the dipole is oriented parallel to the GNP surface, (b) represents the dipole in a perpendicular orientation.}
\label{fig:3}       
\end{figure*}

\begin{figure}
\resizebox{0.5\textwidth}{!}{%
  \includegraphics{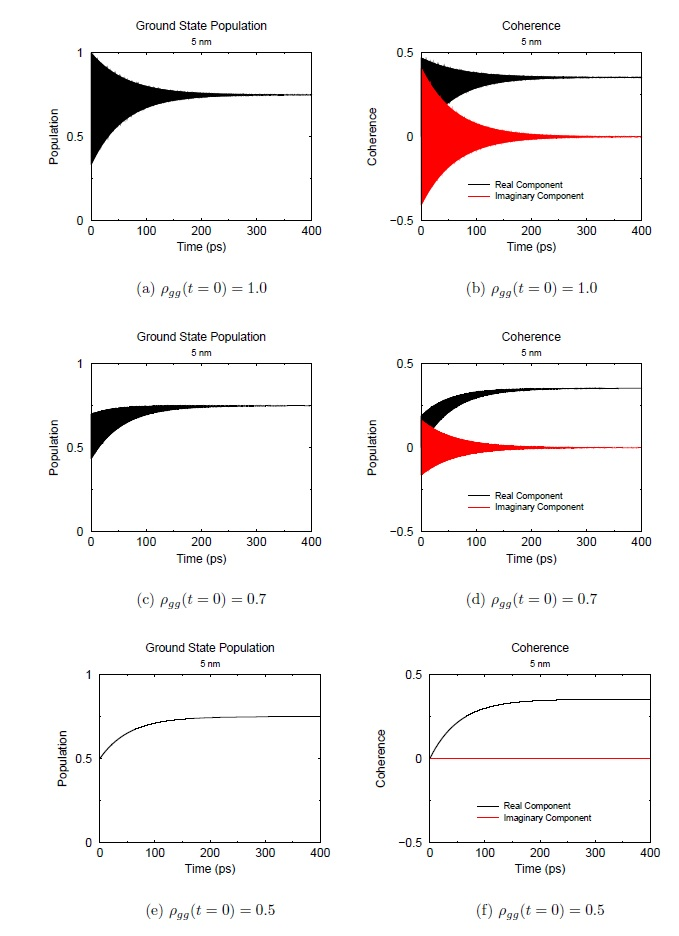}
}
\caption{Ground state populations and coherences of the qubit when placed at distance of 5  nm from the GNP and is evolved from a variety of initial conditions to reach the desired state populations  $\rho_{gg} = 0.75, \rho_{ee} = 0.25$}
\label{fig:4}       
\end{figure}

\begin{figure*}
 \includegraphics[width=1.0\textwidth]{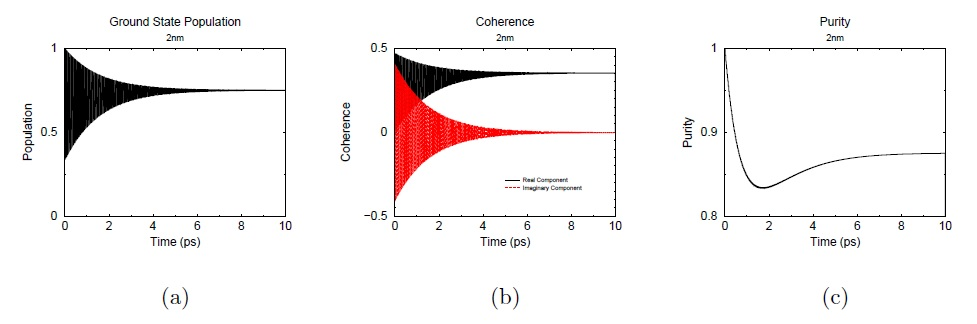}
\vspace*{0.1cm}       
\caption{State dynamics and purity of the qubit when placed at distance of 2  nm from the GNP, this qubit is initially in a pure, ground state and evolves reach the desired state populations  $\rho_{gg} = 0.75, \rho_{ee} = 0.25$}
\label{fig:5}       
\end{figure*}

Moreover, the main benefit of a GNP is to help reduce the time required to reach the steady-state condition by enhancing the decay rate. The time the system takes to reach a final steady state depends on the spontaneous rate of decay experienced by the upper state. When placed near a gold nanoparticle, the time taken to reach the steady state is inversely proportional to the amount of decay enhancement. For the evolution of the state populations  $\dot{\rho_{ee}}=i*\Omega_{ge}/2( \rho_{eg}- \rho_{ge})-\gamma \rho_{gg} $  . Therefore the excited state populations will experience an overall exponential convergence as there exists a term in the evolution equation that is linear with respect to $\gamma$. Therefore an increase in the decay rate ($\gamma$) should linearly correspond to a reduced convergence time due to this timescale. Consider a system placed at a distance of 2nm versus 5nm. At these distances the decay rate is roughly 34 times larger for the 2nm case (figure 5) than the 5nm case (figure 4) and therefore the corresponding convergence time is roughly 34 times larger in figure 5 than in figure 4 as well. Figure 3 shows that, the decay rate enhancement of a qubit placed at a distance of 2 nm from a GNP is approximately $2.4\times10^4$ times of a decay rate of a bare qubit, causing the same order of differences in the needed time to reach the steady state in case of bare qubit.

\begin{figure}
\resizebox{0.5\textwidth}{!}{%
  \includegraphics{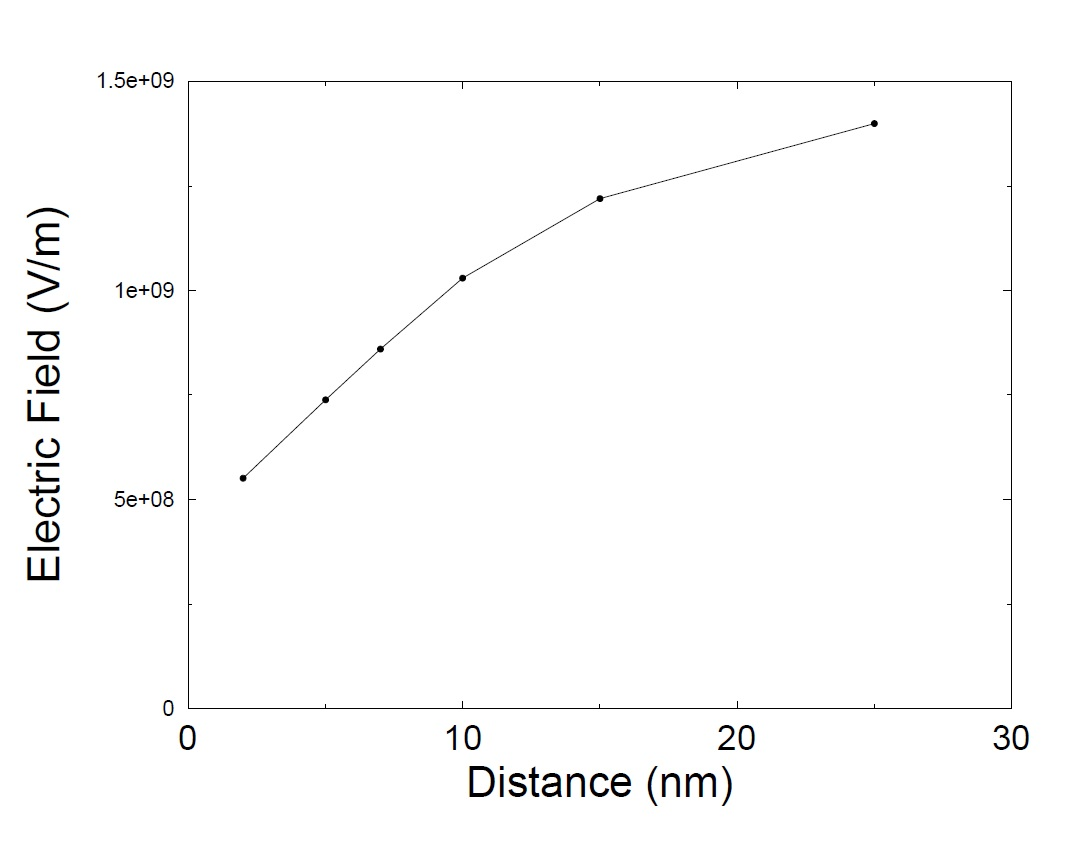}
}
\caption{The required driving electric field versus distance from the surface of GNP}
\label{fig:6}       
\end{figure}

\section{Limitation on Purification in a Two-level System}
\label{sec:1}

For a system with a specific decay rate, enhancement and detuning, choosing the required strength of the electromagnetic field to reach a particular desired ground state population will also set the steady state coherence. As the steady-state purity ($P$) of the two-level system is dependent on only the coherence and ground state population, it will be fixed as well. In fact the steady-state purity of the qubit can be shown, using Equations 5 and 6, to be entirely dependant on the choice of a desired ground state population$\rho_{gg}$ with:
\begin{equation}
	Purity = 4\rho_{gg}-2\rho_{gg}^2-1;
	\end{equation} 

Under this relationship  it is not possible to reach a completely pure state that is not in the ground state configuration using only spontaneous decay and a two-level system.  This limitation restricts this method of state preparation to those in which high levels of purity are not required.

Define the purification factor as $F(P)=P_{output}/P_{input}$ is beneficial to show how effective is this method to prepare a state while improving its purity.  As an example, consider the initial mixed state of $\rho_{gg}=0.75$  and $\rho_{ee}=0.25$ with purity of about 0.625 before going through the process proximate to a GNP. The purity of the output state after the preparation process in presense of a GNP is about 0.875, which is giving F(P)=1.4. Result of purification factors for different initial state purities are presented in figure 7. It can be seen that the $F(P)$ corresponding to this method is always greater than one and just in case that $F_{input} \approx1$ or the intial system is in a totally mixed state, $F(P)$ approaches one.

\begin{figure}
\resizebox{0.5\textwidth}{!}{%
  \includegraphics{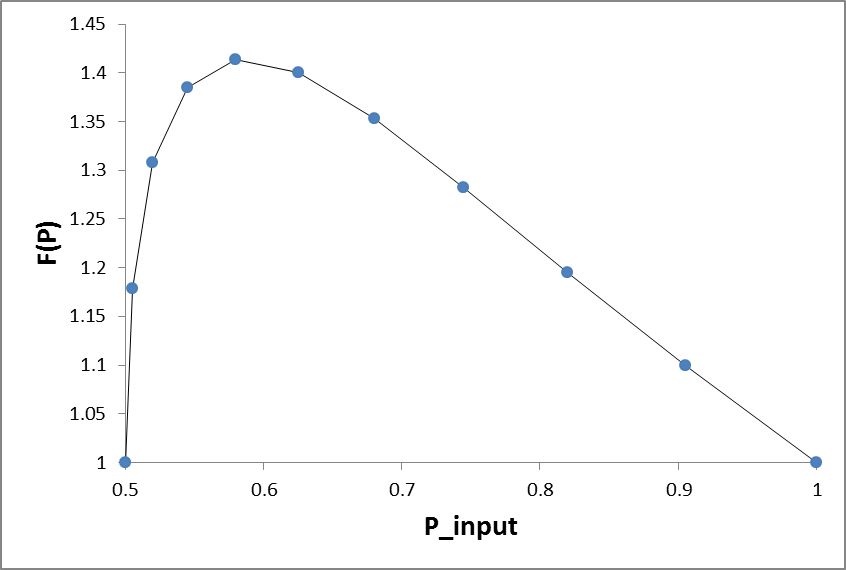}
}
\caption{Results of purification of a qubit }
\label{fig:7}       
\end{figure}

\section{Conclusion}
\label{sec:1}
In this work we show that an incident electromagnetic wave can be used to prepare a qubit in a specific state, regardless of the initial conditions of the qubit. The presence of a proximate GNP can be used to decrease both the required intensity of the incident field as well as the time needed to produce a desired population distribution in a qubit. This process can also be used to cool qubits and increase their state purity in a reduced timescale. The presence of a GNP accomplished this by enhancing both the local electric field around it as well as enhancing the decay rates of the desired states. This reduction in required time, electric field intensity and the ability to modify system purity may represent an improvement in the practical control and use of quantum information systems. 
%

%

%
%
\section{Authors contributions}
All the authors were involved in the preparation of the manuscript.
All the authors have read and approved the final manuscript.
\section {Acknowledgements}
The authors gratefully acknowledge support from the Discovery Grant program of the Natural Sciences and Engineering
Research Council of Canada (Grant No. 311880). Calculations were performed on the SharcNet supercomputing platform.  
We thank Chris DiLoreto for helpful comments and suggestions.

%
%

\end{document}